\documentclass[aps, onecolumn, 11pt]{revtex4}

\usepackage{graphicx}  
\usepackage{subfigure}
\usepackage{amsmath, amsfonts}

\usepackage{fancyhdr}
\usepackage{parskip}
\usepackage[T1]{fontenc}
\usepackage{dcolumn}   

\usepackage{bm}       
\usepackage{amsfonts}  
\usepackage{amsmath}   
\usepackage{amssymb} 
\usepackage{marvosym}
\usepackage{physics}

\newcommand{\pwisein}{\left\{ \begin{array}{ll}}
\newcommand{\pwiseout}{\end{array}\right.}

\numberwithin{equation}{section}

\begin{document}

\title{Higher Order Corrections to Antisymmetric Spherical Solutions in $f(R)$ gravity}

\author{Garrett Gould}
\thanks{Electronic address: gdgould@ucsb.edu}
\affiliation{Department of Physics, University of California, Santa Barbara}

\date{\today}

\begin{abstract}
The static, spherical solutions that exhibit an antisymmetry between the temporal and radial coordinates in $f(R)$ gravity theories are presented. I present the constraint for this antisymmetry and show that pure $R^2$ models produce these solutions. An asymptotic expansion approach is taken to find the solutions, presenting first and second order corrections to the Schwarzschild metric. The null geodesics and photon sphere are determined for these corrections. Furthermore, I show that electromagnetic fields produce this antisymmetry as well. The metric solutions are analyzed in the context of black holes, bringing forward potentially a new class of black holes in modified gravity theories. 
\end{abstract}

\maketitle
\section{Introduction}
Despite the high success of general relativity since its inception, it is accepted to be a low energy effective theory \cite{PhysRevD.50.3874, burgess_2004}, and either lacks detailed description or fails to describe select gravitational phenomena, such as dark energy \cite{RevModPhys.75.559}, cosmic inflation \cite{tsujikawa_2003, universe2040023, PhysRevD.61.023502}, and dark matter \cite{srivastava_immirzi_swain_panella_pacetti_2023, capozziello_cardone_troisi_2006, arun_gudennavar_sivaram_2017, srednicki_2000}. This issue suggests work leading towards describing the higher energy regime of gravitational interactions, reaching a theory of strong gravity. Previous work in modified gravity theories, such as $f(R)$ gravity theory, show that these said phenomena can be described \cite{PhysRevD.70.043528}. In this paper, I analyze the static, spherical solutions in $f(R)$ gravity theories, and pay special attention to those models that have an antisymmetry in the metric between the temporal and radial coordinates, an antisymmetry that is present in the Schwarzschild solution. 

\subsection{$f(R)$ gravity}
The Einstein-Hilbert action, which gives rise to the Einstein field equations, is
\begin{equation}
    S = \frac{1}{16\pi}\int \dd^4 x\; \sqrt{-g}R
\end{equation}
The lagrangian for this action can be generalized to take on any arbitrary function of $R$, $f(R)$. In the metric formalism of $f(R)$ gravity, the action is \cite{fR, faraoni_2008},
\begin{equation}
    S = \frac{1}{16\pi} \int \dd^4 x\; \sqrt{-g}f(R)
\end{equation}
Minimizing the action gives the $f(R)$ field equations,
\begin{equation}
    f'(R)R_{\mu\nu} - \frac{1}{2}f(R)g_{\mu\nu} + (g_{\mu\nu}\Box - \nabla_{\mu}\nabla_{\nu})f'(R) = 8\pi T_{\mu\nu}
\end{equation}
It is useful when determining the dynamics of the equations and solving for the metric tensor by examining the trace of the field equations. For $f(R)$ models, this is,
\begin{equation}
    Rf'(R) - 2f(R) + 3\Box f'(R) = 8\pi T
\end{equation}

\section{Vacuum Spherical Solutions in $f(R)$ gravity}
The general form of a static, spherical metric is \cite{carroll_2022}
\begin{equation}
    \dd s^2 = -e^{2a(r)}\dd t^2 + e^{2b(r)}\dd r^2 + r^2\dd\Omega^2
\end{equation}
Upon which, for the general relativistic case, the unique vacuum solution is the Schwarzschild metric, solved by both $R = 0$ and $R_{\mu\nu} = 0$ \cite{wald_2009}. There exists a symmetry between $R_{tt}$ and $R_{rr}$, upon which for $R = 0$, then $a(r) = -b(r)$, giving rise to anti-symmetry between the temporal and radial coordinates \cite{carroll_2022}. \\
In the simplest nontrivial $f(R)$ model, where $f(R) = R^2$, this anti-symmetry is present. In general, for any choice of $f(R)$, the Ricci tensor in vacuum is,
\begin{equation}
    R_{\mu\nu} = \frac{1}{f'(R)}\left(\frac{1}{2}f(R)g_{\mu\nu} + \nabla_{\mu}\nabla_{\nu} - g_{\mu\nu}\Box f'(R)\right)
\end{equation}
As with the general relativistic case, the setup for determining that the temporal and radial components of the Ricci tensor vanish independently involves setting $e^{2(a - b)}R_{rr} + R_{tt} = 0$ \cite{carroll_2022}, which is true in vacuum for $f(R) = R$. In general, this expression in vacuum is,
\begin{equation}
    e^{2(a - b)}R_{rr} + R_{tt} = \frac{1}{f'(R)}\nabla_r\nabla_rf'(R)
\end{equation}
Where all non-radial derivatives of $f'(R)$ go to zero due to the scalar curvature being a single variable function of the radial coordinate. For any static, spherical metric in $f(R)$ gravity, the scalar curvature must be time-independent, a result guaranteed by the extension of Birkhoff's theorem to $f(R)$ gravity \cite{Birkhoff_1, Birkhoff_2}. \\
In order for the symmetry between the temporal and radial metric coefficients to present itself in $f(R)$ gravity models, the right hand side of equation (2.3) must go to zero. 
\begin{align}
    \frac{1}{f'(R)}\nabla_r\nabla_rf'(R) &= \frac{1}{f'(R)}\left(\partial_r^2 f'(R) + \Gamma_{rr}^{r}\partial_rf'(R)\right) \\
    &= \frac{1}{f'(R)}\left(\partial_r^2 f'(R) + \partial_rb \partial_rf'(R)\right)
\end{align}
So, for any case of $f(R)$ such that $\partial_r^2 f'(R) + \partial_rb\partial_rf'(R) = 0$, the metric will have $a(r) = - b(r)$. 
\subsection{Pure $R^2$ solutions}
Consider the simplest nontrivial choice of $f(R)$, which is $f(R) = \alpha R^2$, where $\alpha$ is a normalization constant. I will work in units such that $\alpha$ = 1 to simplify all equations. The trace equation for this model simplifies down to,
\begin{equation}
    \Box R = 0
\end{equation}
Expanding the d'Alembert operator for this metric, 
\begin{equation}
    \Box R = g^{\mu\nu}\nabla_{\mu}\nabla_{\nu} = g^{rr}\nabla_{r}\nabla_{r}
\end{equation}
Where $\Box R$ is reduced to derivatives of the radial coordinate due to the scalar curvature being time-independent and spherically symmetric. Since this equation is zero, then the $R^2$ model in vacuum must satisfy the condition in equation (2.5), and therefore the metric will be anti-symmetric between the temporal and radial coordinate. \\
In this model, even in vacuum, there should exist non-zero values of $R$ due to the second-order homogeneous equation in (2.7). This can also be seen by solving for the Ricci tensor in (1.3). 
\begin{equation}
    R_{\mu\nu} = \frac{1}{f'(R)}\left(\frac{1}{2}f(R)g_{\mu\nu} + \nabla_{\mu}\nabla_{\nu}f'(R) - g_{\mu\nu}\Box f'(R)\right)
\end{equation}
So in the case of $f'(R) \neq 1$, the Ricci tensor will have non-zero values in vacuum.
Because of the anti-symmetry between $R_{tt}$ and $R_{rr}$, it will be best to solve for $R$ by solving both the trace equation (2.6) and solving for the angular Ricci tensor,
\begin{equation}
    R_{\theta\theta} = \frac{1}{4}r^2 R
\end{equation}
Where $\Box R = 0$ was substituted to simplify the equation. The primary aim of this section is to find the metric for this specific model. Rewriting the metric in (2.1) such that $e^{2a(r)} = A(r)$,
\begin{equation}
    \dd s^2 = -A(r)\dd t^2 + \frac{1}{A(r)}\dd r^2 + r^2\dd \Omega^2
\end{equation}
In terms of the metric coefficient $A(r)$, the scalar curvature $R$ is \cite{carroll_2022},
\begin{equation}
    R(r) = -\frac{2}{r^2}\left(A(r) + 2rA'(r) - 1\right) - A''(r)
\end{equation}
Upon which the homogeneous solution to this equation, as well as the angular Ricci tensor equation is the Schwarzschild metric. So, for non-zero scalar curvature values, there should exist a metric such that,
\begin{equation}
    A(r) = 1 - \frac{2M}{r} + \epsilon(r)
\end{equation}
Where $\epsilon(r)$ is the inhomogeneous coupled solution between the angular Ricci tensor and the scalar curvature. This extra term can be considered to be the "correction" to the Schwarzschild metric arising from considering a modified gravity theory such as $f(R) = R^2$. 
In terms of the metric in (2.10), the $R_{\theta\theta}$ component is \cite{carroll_2022},
\begin{align}
    R_{\theta\theta} &= 1 - A(r) - rA'(R) \\
    &= -\epsilon(r) - r\epsilon'(r)
\end{align}
Setting (2.9) and (2.14) equal to each other, I get the second differential equation to solve for this case, 
\begin{equation}
    -\epsilon(r) - r\epsilon'(r) = \frac{1}{4}r^2 R(r)
\end{equation}
Since the highest order term of $R(r)$ in terms of the metric will come from $A''(r)$, and $R(r)$ vanishes for $A(r) \sim 1/r^2$, then I can establish a constraint on the possible choices of $R(r)$,
\begin{equation}
    \lim_{r \rightarrow \infty} R(r) r^2 = 0
\end{equation}
A constraint that is required for all asymptotically flat spacetimes, and also shown in \cite{pretel_duarte_2022}. Since $R(r)$ must now be asymptotically flat and approach flatness faster than the growth of $r^2$, a possible solution of $R(r)$ is an asymptotic expansion about $r = 0$, 
\begin{equation}
    R(r) = \sum_{n = 2}^{\infty}\frac{b_n(2M)^{n-2}}{r^n}
\end{equation}
Where powers of the Schwarzschild radius, $2M$, are included to preserve units. From (2.3) and (2.11), a choice for the perturbation to the metric is another asymptotic expansion about $r = 0$, 
\begin{equation}
    \epsilon(r) = \sum_{n = 0}^{\infty}\frac{a_n(2M)^n}{r^n}
\end{equation}
Now, to solve for the coefficients of the $R$ expansion, I solve (2.7) for a static, spherical metric. This equation becomes,
\begin{equation}
    A\partial_r^2 R + \frac{2}{r}\partial_r R - \partial_rA\partial_rR = 0
\end{equation}
Because of the highly non-linear nature of equation (2.15), a series solution approach will provide a great approximation to the solutions, especially an asymptotic series approach. Direct substitution of equation (2.13), (2.14) into the differential equation (2.15) gives,
\begin{align}
    \sum_{n = 2}^{\infty}\frac{n(n-1)b_n(2M)^{n-2}}{r^{n+2}} - \sum_{n = 2}^{\infty}\frac{n^2b_n(2M)^{n-1}}{r^{n+3}} + \sum_{n=2}^{\infty}\frac{(2M)^{n-4}}{r^n}\left(S_{n}^{(1)} - S_{n}^{(2)}\right) = 0
\end{align}
Where,
\begin{align}
    S_{n}^{(1)} &= \sum_{l = 0}^{n}(n-l)(n-l+1)a_{l-2}b_{n-l} \\
    S_{n}^{(2)} &= \sum_{l = 0}^{n}(l-2)(n-l)a_{l-2}b_{n-l}
\end{align}
To account for the nonlinear terms of the differential equation, I employed the Cauchy product (cite here), which gives rise to the finite sums in equations (2.17), (2.18). Since the powers and the starting index of the sums are not aligned, the first few terms of the first and third infinite series in (2.16) need to be extracted and evaluated. This gives the three equations,
\begin{align}
    S_2^{(1)} - S_{2}^{(2)} &= 0 \\
    S_3^{(1)} - S_{3}^{(2)} &= 0 \\
    S_4^{(1)} - S_{4}^{(2)} + 2 b_2 &= 0
\end{align}
Expanding out these equations,
\begin{align}
    3a_{-1}b_1 + 10a_{-2}b_2 = 0 \\
    2a_0b_1 + 8a_{-1}b_2 + 18a_{-2}b_3 = 0 \\
    a_1b_1 + 2b_2 + 6a_0b_2 + 15a_{-1}b_3 + 28a_{-2}b_4 = 0
\end{align}
Examining the first equation, since $a_{-1} = a_{-2} = 0$, this is statement that the expression is true, and $b_2$ is undefined from this equation. Moving on to the second equation, $b_2$ and $b_3$ are also undefined, as well as $a_0$. So, the third equation will be needed to determine these coefficients. Solving for $b_2$ gives,
\begin{equation}
    b_2 = \frac{-a_1b_1 - 15a_{-1}b_3 - 28a_{-2}b_4}{2+6a_0}
\end{equation}
Upon which $b_2 = 0$. Now, since the $a_0$ term in the expansion of $\epsilon(r)$ will be a constant, and will not contain any powers of $r$, then it is assumed that $a_0 = 0$, otherwise the constant term of the metric will be not equal to one, which is a requirement for the metric to be asymptotically flat. Since this is a constant, a transformation of the coordinates could also take care of this problem, but I can set $a_0 = 0$ without loss of generality. It will be shown later that this is concluded as well when relating $b_n$ to $a_n$ via the $R_{\theta\theta}$ equation. \\
Since $b_2 = 2$, the recursion relation for higher terms starts for $n > 2$, and is
\begin{equation}
    b_{n+1} = \frac{1}{n(n+1)}(n^2b_n + S_{n+3}^{(2)} - S_{n+3}^{(1)})
\end{equation}
All coefficients in the series can be expressed by the preceding coefficient. Since the non-zero series coefficients start at $b_3$, and this coefficient can take on any numerical value without loss of generality, the series expression of $R$ can be expanded in terms of $b_3$
\begin{align}
    R(r) = \frac{(2M)}{r^3}b_3 - \frac{3}{4}\frac{(2M)^2}{r^4}b_3(a_1 - 1) + \frac{(2M)^3}{r^5}\left(\frac{3}{5}b_3(a_1 - 1)^2 - \frac{3}{10}a_2b_3\right) + O\left(\frac{1}{r^6}\right) 
\end{align}
Where $a_1, a_2$ are determined by solving for $\epsilon(r)$ using (2.11). 
Taking the first order of $R$ and solving (2.11) gives $a_1$,
\begin{equation}
    a_1 = -\frac{b_3}{4}\ln(r/r_0)
\end{equation}
Where $r_0$ is a length scale parameter introduced when solving the differential equation in order to keep the argument of the logarithm dimensionless. Higher terms of $a_n$ are solved to be,
\begin{equation}
    a_n = \frac{1}{4(n-1)}b_{n+2}, \; \; n > 1
\end{equation}
Upon which, these coefficients give the corrections to the metric. So, as previously stated, the $a_0$ term in the expansion of $\epsilon(r)$ must be zero. Relabeling $b_3 = \xi$, the higher order corrected metric is now,
\begin{align}
    A(r) = 1 - \frac{2M}{r} - \frac{M}{2}\frac{\xi\ln(r/r_0)}{r} 
    + \frac{3}{16}\frac{M^2}{r^2}\left(\xi^2\left(1 + \ln(\frac{r}{r_0})\right) + 4\xi\right) + O\left(\frac{1}{r^3}\right)
\end{align}
Since higher-order corrections to the metric than stated in equation (2.34) will contain higher powers of $\xi$, in order for these additional terms to be considered corrections and not diverge, the constraint $|\xi| < 1$ is needed. Since the logarithm term will grow at a slower rate than the $r^{-n}$ terms, no further constraints are needed on the logarithm terms in order to ensure convergence. 
\section{Vacuum spherical geodesics}
The first application of the results derived in Section 2.1 is the effects of the corrections on the geodesics, specifically the corrections to the stable orbits. In general, the geodesics will be determined by \cite{carroll_2022},
\begin{equation}
    \dv[2]{x^{\mu}}{\tau} + \Gamma_{\lambda\nu}^{\mu}\dv{x^{\lambda}}{\tau}\dv{x^{\nu}}{\tau}
\end{equation}
For any metric coefficient $A(r)$, the orbit potential will be,
\begin{equation}
    V(r) = \frac{1}{2}A(r)\left(\frac{L^2}{r^2} + \kappa\right)
\end{equation}
Where $\kappa = -g_{\mu\nu}U_{\mu}U_{\nu}$ is the normalization constant of the four-velocity. In the case of null geodesics, this will always be zero. \\
To first order, the potential in the $R^2$ model is,
\begin{equation}
    V(r) = \frac{1}{2}\left(1 - \frac{2M}{r} - \frac{M}{2}\frac{\xi \ln(r/r_0)}{r}\right)\left(\frac{L^2}{r^2} + \kappa\right)
\end{equation}
\begin{figure}[h!]
    \centering
    \includegraphics[scale=0.667]{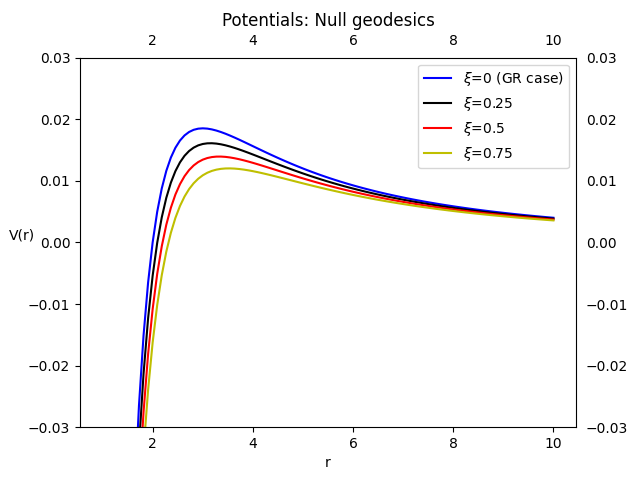}
    \caption{The potentials for massless particles are plotted, varying in $\xi$. The maxima decrease with $\xi$, but the location of maxima, where the stable orbital radii are, increase.}
    \label{fig:nul_pot}
\end{figure}
\\The stable orbital radii can be determined by minimizing the potential. In the case of massless particles, where $\kappa = 0$, 
\begin{equation}
    r_c = 3M\left(1 - \frac{\xi}{12} + \frac{\xi \ln(r_c/r_0)}{4}\right)
\end{equation}
Because of the logarithm term, the solution will involve the Lambert W function \cite{corless_gonnet_hare_jeffrey_knuth_1996}. The photon sphere radius in the first order corrected metric is,
\begin{equation}
    r_c = 3M - \frac{3}{4}M\xi W_0(z)
\end{equation}
Where $W_0$ is the principal branch of the Lambert W function, and
\begin{equation}
    z = -\frac{4r_0}{3M\xi}\exp(\frac{1}{3} - \frac{4}{\xi})
\end{equation}
\begin{figure}[h!]
    \centering
    \includegraphics[scale=0.667]{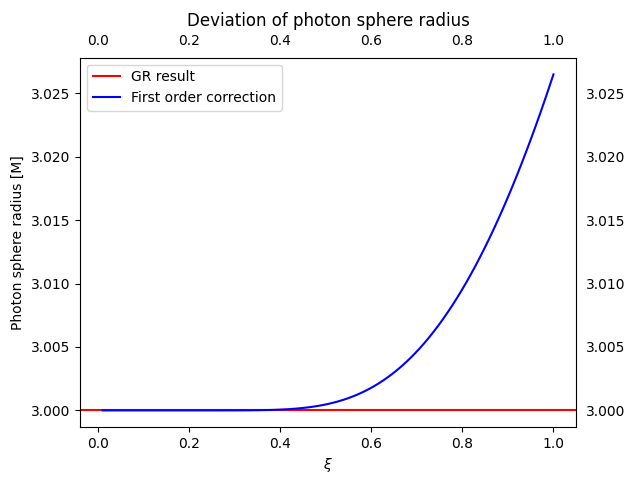}
    \caption{The general relativistic and the first order corrected photon sphere radii are plotted. The photon sphere due to the first order corrections is increased slightly as a function of $\xi$. This corresponds with \ref{fig:nul_pot}, where the maxima of the potentials shifted towards larger $r$ at larger values of $\xi$. }
    \label{fig:photon_sphere_first}
\end{figure} 
\\ Because of the logarithm term, all higher order corrections require numerical evaluations of the photon sphere radius, as there will not be a closed form solution. 
\section{$f(R)$ spherical solutions with matter sources}
When introducing a matter field, the Ricci tensor in (2.2) becomes,
\begin{equation}
    R_{\mu\nu} = \frac{1}{f'(R)}\left(8\pi T_{\mu\nu} + \frac{1}{2}f(R)g_{\mu\nu} + \nabla_{\mu}\nabla_{\nu}f'(R) - g_{\mu\nu}\Box f'(R)\right)
\end{equation}
So, for $R^2$ models, the trace equation is,
\begin{equation}
    \Box R = \frac{8\pi}{3}T
\end{equation}
And the angular Ricci tensor component is,
\begin{equation}
    R_{\theta\theta} = \frac{1}{2R}\left(8\pi T_{\theta\theta} + \frac{1}{2}r^2 R^2 - \frac{8\pi}{3}r^2T\right)
\end{equation}
For the anti-symmetry between the temporal and radial components of the metric to exist when introducing a source, the equation (2.3) is generalized to include source terms,
\begin{equation}
    e^{2(a - b)}R_{rr} + R_{tt} = \frac{1}{f'(R)}\nabla_{r}\nabla_{r}f'(R) + \frac{8\pi}{f'(R)}(g_{tt}g^{rr}T_{rr} + T_{tt})
\end{equation}
So, for this anti-symmetry to continue over when introducing a matter field, there needs to be the same anti-symmetry in the energy-momentum tensor, that is,
\begin{equation}
    g^{rr}T_{rr} = -g^{tt}T_{tt}
\end{equation} 
Matter fields that automatically satisfy this relation are those with a traceless energy-momentum tensor, such as for electromagnetic fields. 
\subsection{$R^2$ corrections with electromagnetic fields}
Since the trace of the electromagnetic energy-momentum tensor, then the trace equation for $R^2$ remains the same as vacuum, and therefore $R(r)$ is unchanged. But, the metric will gain terms corresponding to the electric and magnetic fields. This is a similar process to that of the Reissner-Nordstr\"{o}m metric, where the electric field is introduced as a second term to the Schwarzschild metric \cite{wald_2009, carroll_2022}. \\
To solve for the metric, I employ equation (2.15), but with a source term introduced,
\begin{equation}
    -\epsilon(r) - r\epsilon'(r) = \frac{1}{4}r^2 R + \frac{4\pi}{R}T_{\theta\theta}
\end{equation}
The angular component of the electromagnetic energy-momentum tensor is \cite{PhysRevD.55.809, ANGHINONI2022169004},
\begin{align}
    T_{\theta\theta} &= -r^4\sigma_{\theta\theta} \\
    &= \frac{1}{2}\frac{Q_0^2}{16\pi^2}
\end{align}
Where $\sigma_{ij}$ is the Maxwell stress tensor, and $Q_0^2 = (Q_e^2 + Q_m^2)$ represents the total electric and magnetic charge of the system. Although $R(r)$ exhibits the same solutions in the case of an electromagnetic field source, I transform the powers of $2M$ in the expansion of the metric and the scalar curvature to a general variable $\mathcal{R}$, upon which the coordinate singularity radius for each specific case would be substituted. In the static case, $\mathcal{R} = 2M$, and for the non-rotating charged case, $\mathcal{R} = \frac{1}{2}\left(2M \pm \sqrt{4M^2 - Q_0^2/\pi}\right)$. So, the expansion of $R(r)$ becomes, 
\begin{equation}
    R(r) = \frac{\mathcal{R}}{r^3}\xi - \frac{3}{4}\frac{\mathcal{R}^2}{r^4}\xi(a_1 - 1) + \frac{\mathcal{R}^3}{r^5}\left(\frac{3}{5}\xi(a_1 - 1)^2 - \frac{3}{10}a_2\xi\right) + O\left(\frac{1}{r^6}\right) 
\end{equation}
For the case of the electromagnetic field source, the $a_1$ coefficient of the metric expansion is,
\begin{equation}
    a_1 = -\frac{1}{4}\mathcal{R}\xi\ln(r/r_0) - \frac{Q^2(r)}{32\pi\xi\mathcal{R}^2}
\end{equation}
Where $Q(r) = Q_0r^2$. The metric is then, 
\begin{equation}
    A(r) = 1 - \frac{2M}{r} - \frac{M}{2}\mathcal{R}\xi\frac{\ln(r/r_0)}{r} - \frac{M Q^2(r)}{16\pi\xi\mathcal{R}^2 r} + O\left(\frac{1}{r^2}\right)
\end{equation}
Because $Q(r)$ will overpower the denominator at large $r$, the metric will not be asymptotically flat. This is due to the $1/R$ term in the angular Ricci tensor equation. A power series solution for $R(r)$ would overcome this, but then would be significantly overpowered by the first term on the right-hand-side of the angular Ricci tensor equation. So, for an antisymmetric metric with an electromagnetic source, the spacetime would be highly unstable, rendering the $R^2$ model unphysical in the case of an antisymmetric metric. 
\section{Modified Black Holes}
Even though the electromagnetic case was rendered to be unphysical for the specific metric considered in this paper, a point mass $M$ can still be considered. Because of the corrections to the metric in equation (2.34), the dynamics of the event horizons and the coordinate singularities will change, in comparison to the Schwarzschild metric. Consider the first order corrections to the metric,
\begin{equation}
    A(r) = 1 - \frac{2M}{r} - \frac{M}{2}\frac{\xi\ln(r/r_0)}{r}
\end{equation}
There are two things to note here. First, when $r = r_0$, the classical Schwarzschild metric is produced, regardless of the choice of $\xi$. Second, the slope of the light cone behaves differently than in the general relativistic case. The slope is \cite{carroll_2022},
\begin{equation}
    \frac{dt}{dr} = \pm \left(1 - \frac{2M}{r} - \frac{M}{2}\frac{\xi\ln(r/r_0)}{r}\right)^{-1}
\end{equation}
Which diverges not only at a new coordinate singularity location larger than the Schwarzschild radius, depending on the value of $\xi$. But, the slope of the light cone flips direction again once reaching an inflection point at the maximum point of the slope. Transforming to a modified version of Eddington-Finkelstein coordinates for infalling geodesics to represent this behavior \cite{PhysRev.110.965}, the metric is,
\begin{equation}
    \dd s^2 = -\left(1 - \frac{2M}{r} - \frac{M}{2}\frac{\xi\ln(r/r_0)}{r}\right)\dd v^2 + \dd v\dd r +\dd r \dd v + r^2\dd \Omega^2
\end{equation}
The maximum of the slope of the light cone occurs at,
\begin{equation}
    r_{\text{max}} = r_0 \exp(1 - \frac{4}{\xi})
\end{equation}
At this point, the light cone slope rapidly increases, until crossing a second coordinate singularity that arises from the first order correction. Upon crossing this point, the light cone inverts its behavior back to when it was outside the first coordinate singularity, with a rapidly decreasing slope. The second coordinate singularity occurs at,
\begin{equation}
    r = -\frac{1}{2}M\xi W_0\left(z_{\text{EH}}\right)
\end{equation}
Where the argument of the Lambert W function is,
\begin{equation}
    z_{\text{EH}} = - \frac{2r_0}{M\xi}e^{-4/\xi}
\end{equation} \\ \\ \\ 
\begin{figure}[h!]
    \centering
    \includegraphics[scale=0.667]{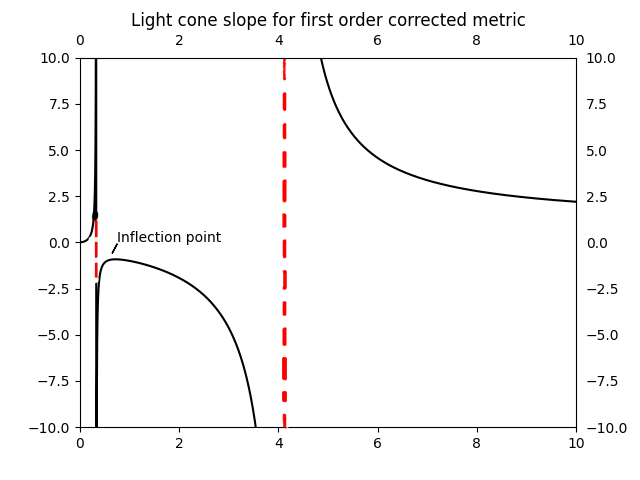}
    \caption{Plot of the slope of the light cone. To show the inflection point. As with previous figures, the units are such that $M = r_0 = 1$. To emphasize the location of the inflection point, $\xi = 3$, even though it is constrained to $|\xi| < 1$ from Section 1. The two coordinate singularities are highlighted as red dashed lines.}
    \label{fig:light cone first order}
\end{figure}
\begin{figure}[h!]
    \centering
    \includegraphics[scale=0.333]{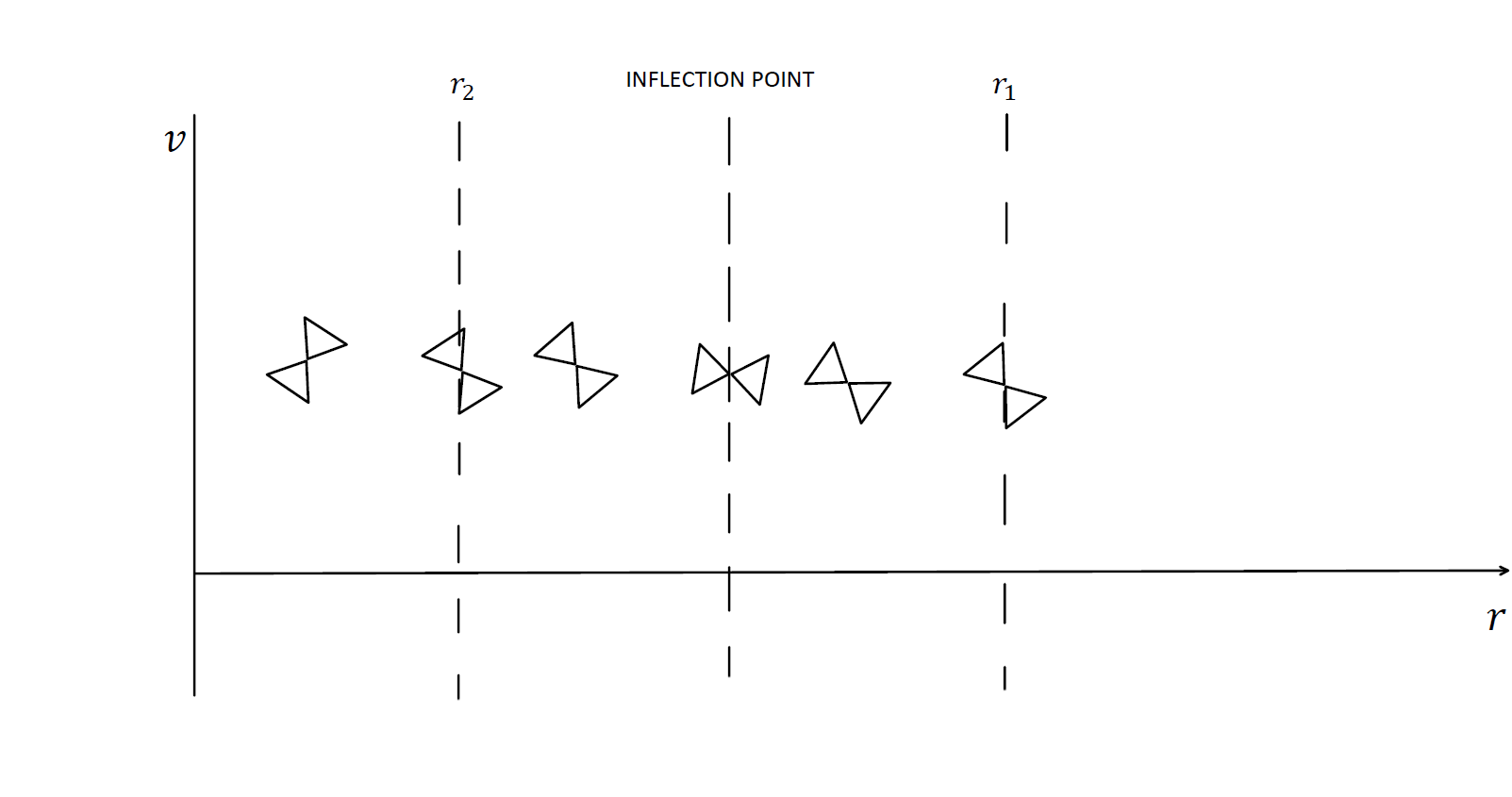}
    \caption{Diagram of the light cones when crossing the two coordinate singularities $r_1$ and $r_2$, as well as the inflection point. The diagram is on a v-r plot, indicating infalling geodesics.}
    \label{light cone diagram}
\end{figure}

\section{Conclusion}
The static, spherical metric solutions in $f(R) = R^2$ that exhibit an antisymmetry between the temporal and radial coordinates were presented by an asymptotic series expansion due to the singularity at $r = 0$. To ensure each higher order term does not rapidly diverge, a constraint on the expansion coefficient $\xi$ is set such that $|\xi| < 1$. Because of this constraint, all deviations away from general relativity due to the corrections from $R^2$ gravity are small. This presents challenges when testing for these corrections. For example, at the limit of the $\xi$ constraint, the photon sphere radius increases by $M/40$, a very small deviation away from general relativity. \\
\indent When introducing a matter source, the only viable energy-momentum tensor that would conserve the antisymmetric metric was an electromagnetic field, due to the traceless nature of its energy-momentum tensor. But, after further analysis, these results and the corrections that the electromagnetic fields produced onto the metric in $R^2$ gravity were shown to either be unphysical or highly unstable, indicating that a different approach to finding the scalar curvature and metric in the presence of an electromagnetic field is needed. The condition of the antisymmetry between the radial and temporal coordinate may need to be relaxed in order to determine these solutions. 
\indent In the case of black holes, in the first order correction, a second horizon presented itself. With the constraint on $\xi$, this second horizon would be very close to the curvature singularity at the center. There exists a point of inflection, where the tilting of the light cones due to the crossing of the first event horizon reverts, and the slopes of the light cone rapidly increases as the test particle falls towards the second horizon. Upon crossing the second horizon, the light cone points outwards, behaving inversely to the outer event horizon, preventing any infalling matter from reaching the center. \\
\indent It is possible that this second horizon provides a "shell" around the curvature singularity at $r = 0$. Test particles that fall inwards inside the black hole would be trapped, since they would have a strong gravitational field acting inwards towards the center, but upon crossing the second horizon, the test particles would experience an inverse force acting outwards, launching the particles outside of the second horizon, trapping them around this region. This would create a shell of matter that surrounds the singularity, never reaching it, and only space-like geodesics crossing the second horizon could reach the singularity, inversely to the first event horizon.  \\
\indent Taking this into consideration, a new class of black holes, specifically black holes in modified gravity theories like $f(R)$ gravity, may be imminent. Further development of the solutions in $f(R)$ gravity models, such as other choices of models, and other modified theories of gravity are needed. Such an example is $f(R, T)$ gravity, where the energy-momentum trace is directly included into the Einstein-Hilbert action \cite{fRT}. The further study of spherical solutions, and potentially other rotating models such as Kerr-Newman black holes, may be a route in finding further deviations away from general relativity that present themselves in modified gravity theories. These deviations are likely a great route in testing and falsifying these models of gravity to determine what is needed to seek for a effective theory of gravity that provides better insight than general relativity. 
\bibliography{references}

\end{document}